\documentclass{article}

\usepackage{PRIMEarxiv}

\usepackage[utf8]{inputenc} 
\usepackage[T1]{fontenc}    
\usepackage{hyperref}       
\usepackage{url}            
\usepackage{booktabs}       
\usepackage{amsfonts}       
\usepackage{nicefrac}       
\usepackage{microtype}      
\usepackage{lipsum}
\usepackage{fancyhdr}       
\usepackage{graphicx}       
\graphicspath{{media/}}     

\title{The Interactive Modeling of a Binary Star System

}

\author{
  Canberk Soytekin \\
  Çakabey Schools \\
  Izmir, Turkey\\
  \texttt{canberk.soytekin@cakabeyli.com} \\
   \And
  Ayşe Pelin Dedeler \\
  Çakabey Schools \\
  Izmir, Turkey\\
  \texttt{ayse.pelin.dedeler@cakabeyli.com} \\
}

\begin{document}
\maketitle

\begin{abstract}
An interactive binary star system simulation was developed to be showcased on an educational platform. The main purpose of the project is to provide insight into the orbital mechanics of such star systems with the help of a three-body simulation. The initial simulation script was written in the Python programming language with the help of the VPython addition. The custom-made models were created on Blender and exported. For the final implementation of the simulation on the Godot game engine, the Python code was converted into GDScript and the Blender models were re-textured. 

\end{abstract}


\section{Introduction}

The three-body problem was posed by Newton in Principia \cite{newton} , where he considered the motion of the Earth and the Moon around the Sun. Since then, the three-body problem has been one of the most well-known problems in dynamical astronomy \cite{Whittaker}.Extensive theoretical and numerical research has been dedicated to the study of this attractive problem; however, for simplicity's sake, the majority of research has focused on the \textit{circular constrained} problem \cite{deAlmeidaJunior2022}. The development of high-speed computers has increased interest in simulating the general three-body problem. Applications of the three-body problem in astronomy are essential for determining how three stars, a star with a planet that has a moon, or any other set of three celestial objects can maintain a stable orbit. Three-body problem based foundational models of the gravitational influence of the Sun on Earth and the Moon were investigated while constructing significant space missions like the James Webb Space Telescope \cite{webb}. The Hubble Space Telescope, the Chandra X-Ray Observatory, the SPITZER Space Telescope, and the Kepler Space Telescope are some of the most noteworthy discovery missions that NASA has conducted; each of these missions have used a distinct approach to the three-body problem\cite{Musielak}.

 \begin{figure}[h]
  \centering
  \includegraphics[width=10 cm]{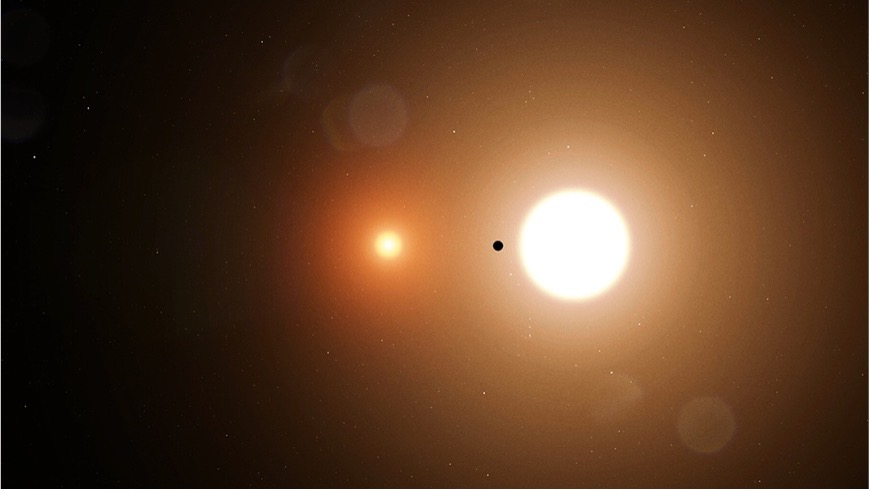}
  \caption{The TOI-1338 Binary Star System \cite{nasa}}
\end{figure}

 Scientists are increasingly favoring active learning tools \cite{learn}. Simulations may not only serve to motivate students, but also to enhance their intuitive knowledge of abstract physics problems.\cite{learn2}, To achieve our goal of turning the three-body problem into an interactable topic for students, researchers, and enthusiasts, we have created a simulated system consisting of two stars and one observatory satellite which are in elliptical orbits determined by a differentially simulated application of classical mechanics principles. The initial parameters of the observatory satellite: position (x, y, z) and velocity (x, y, z) vectors can be given by the user to initiate the simulation.

\section{The Scientific Groundwork}
\label{sec:headings}

The scientific bases of the project were obtained from Orbital Mechanics by Vladimir Chobotov \cite{1}, Elements of Newtonian Mechanics by Jens M. Knudsen and Poul G. Hjorth \cite{2}, and the MIT 8.01SC Classical Mechanics Course. 

\begin{equation}
m_2 \frac{v_0^2}{R_{12}}=G \frac{m_1 m_2}{R_{12}^2}, \quad
\centering
v_0^2=G \frac{m_1}{R_{12}}, \quad v_0=\sqrt{\frac{G m_1}{R_{12}}} ; \quad
\end{equation}

\begin{equation}
\overrightarrow{p_{02}}=\left[\begin{array}{c}
0 \\
\mathrm{~m}_2 \cdot \mathrm{v}_0
\end{array}\right], \quad \sum \vec{p}=0, \quad \overrightarrow{p_{01}}-\overrightarrow{p_{02}}
\end{equation}

Conservation of momentum and Kepler’s First Law for celestial bodies were utilized in determining the initial velocities of the binary stars as well as satisfying the orbital prerequisites.

\begin{equation}
\overrightarrow{r_{H 0}}=R \textsubscript{12}\left[\begin{array}{l}
r_x \\
r_y \\
r_z
\end{array}\right], \quad \overrightarrow{v_{H 0}}=\mathrm{v}_f\left[\begin{array}{l}
v_x \\
v_y \\
v_z
\end{array}\right]
\end{equation}

The scalar properties of the satellite are mass (m\textsubscript{0}) and initial speed (v\textsubscript{f}). The vectoral properties are its position vector and the components of its velocity.

 \begin{figure}[h]
  \centering
  \includegraphics[width=7.5 cm]{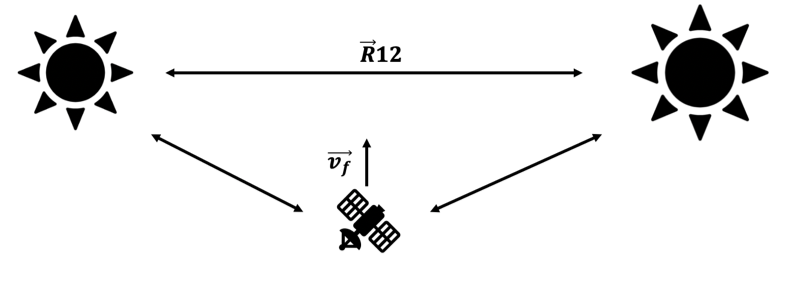}
  \caption{The diagram of the initial conditions.}
\end{figure}

\section{Python Simulation}
In the mathematical modeling of the scientific bases of the project, a differential calculation method was utilized.
In the Python simulation, the positions and the momenta of the stars and the satellite were updated approximately every 200 microseconds (dt = 200 \textmu s).

Throughout each dt time interval, the forces on the bodies and their momenta were assumed to be constant.

The momenta were updated differentially according to the following equations:

\begin{equation}
\overrightarrow{p_{new}}=\overrightarrow{p_{old}}+\overrightarrow{\Delta p}, \quad \overrightarrow{\Delta p}=\int_{t_0}^{t_0 + \Delta t} \vec{F} d t
\end{equation}

\begin{equation}
\lim _{\Delta t \rightarrow 0} \quad
\overrightarrow{\Delta p}=\vec{F} \cdot \Delta t, \quad
\vec{p}_{new}^{\prime}=\overrightarrow{p_{old}}+\vec{F} \cdot \Delta t
\end{equation}

The positions were updated differentially according to the following equations:

\begin{equation}
\vec{r}_{new}^{\prime}=\overrightarrow{r_{old}}+\overrightarrow{\Delta r}, \quad \overrightarrow{\Delta r}=\int_{t_0}^{t_0 + \Delta t} \vec{v} d t, \quad \vec{p}=m \vec{v}, \quad \vec{F}=\frac{d \vec{p}}{d t}
\end{equation}

\begin{equation}
\lim _{\Delta t \rightarrow 0} \quad
\overrightarrow{\Delta r}=\frac{\vec{p}}{m} \cdot \Delta t, \quad
\vec{r}_{new}^{\prime}=\overrightarrow{r_{old}}+\frac{\vec{p}}{m} \cdot \Delta t
\end{equation}

 \begin{figure*}[ht!]
            \centering
            \includegraphics[width=.6\textwidth]{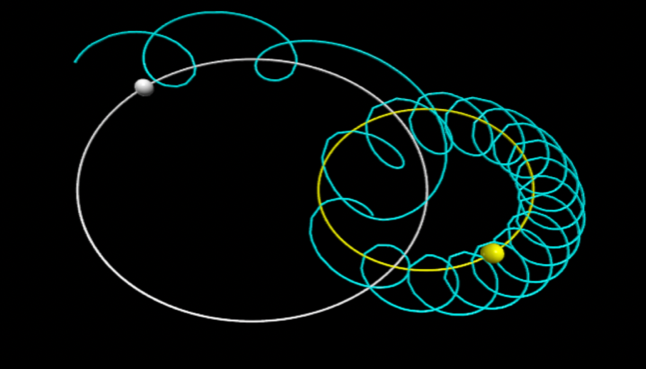}
            \caption{The Python Simulation}
            \centering
            \includegraphics[width=.58\textwidth]{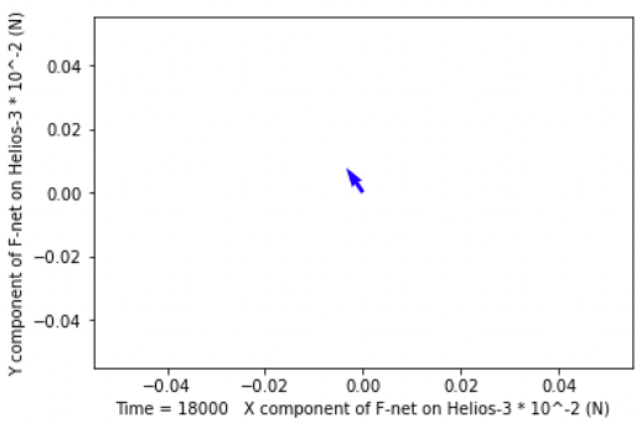}\hfill
            \caption{Time-dependent diagram of the $$
\overrightarrow{F_{n e t}}(t)
$$ vector of the satellite.}
            
        \end{figure*}

The vector analysis and visual Modeling of the bodies were done with the help of the VPython module.

Two varieties of code were written for the Python simulation: a \textit{visual code} and an \textit{analysis code}. In the visual code, the script runs at a rate of 500 (a maximum 500 operations per second, a property of VPython) and is focused on the visual depiction of the simulation; in the analysis code, an intensive vectoral analysis module utilizing numpy and matplotlib was integrated into the while loop in addition to the visualization module. Even through the analysis code runs at 10 times the rate (a maximum of 5000 operations per second) of the visual code, the increased workload causes it to run slower compared to its less intensive counterpart. Both simulation scripts are available on GitHub.

\begin{center}
   \url{https://github.com/KameSanKaro/The-Interactive-Modeling-of-a-Binary-Star-System---Supplementary-Code.git}  
\end{center}

 \begin{figure}[b]
  \centering
  \includegraphics[width=11.99 cm]{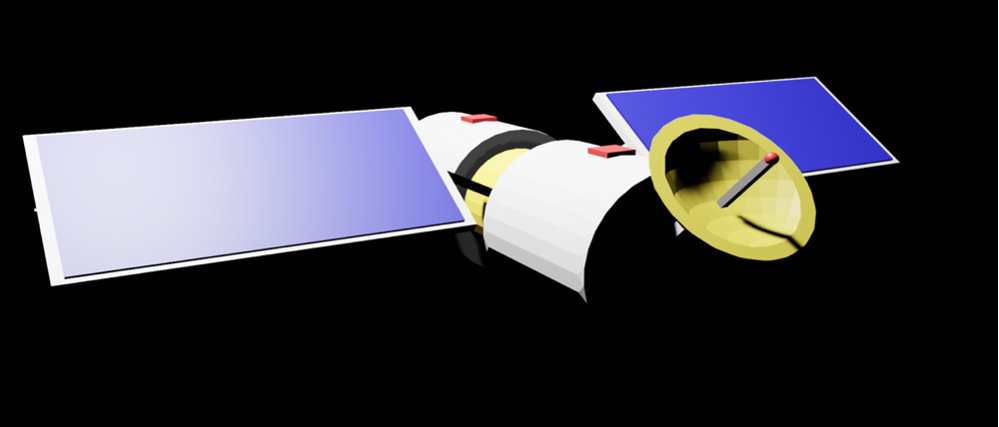}
  \caption{The satellite model.}
\end{figure}

\section{Godot Implementation}

In the Godot implementation of the simulation, the simulation code (visual code) was converted from Python to GDScript for increased Godot compatibility. In addition to the translation of the primary simulation code, various signal scripts were implemented for navigability measures, such as menu screens.

The transfer of variables (initial parameters and the score) between scenes were conducted with the help of the Autoload property of Godot.
The simulation objects were implemented as rigid bodies, and their kinematic and inertial properties were made compatible to those of Godot.
In the Godot implementation process, the code was changed from momentum-based to velocity-based because of Godot not being able to handle the large masses of the objects in the calculations:

\begin{equation}
\overrightarrow{v_{new}}=\overrightarrow{v_{old}}+\overrightarrow{\Delta v}, \quad \overrightarrow{\Delta v}=\int_{t_0}^{t_0 + \Delta t} \vec{a} d t
\end{equation}

\begin{equation}
\lim _{\Delta t \rightarrow 0} \quad
\overrightarrow{\Delta v}=\vec{a} \cdot \Delta t, \quad
\vec{v}_{new}^{\prime}=\overrightarrow{v_{old}}+\vec{a} \cdot \Delta t
\end{equation}

\begin{equation}
\vec{r}_{new}^{\prime}=\overrightarrow{r_{old}}+\overrightarrow{\Delta r}, \quad \overrightarrow{\Delta r}=\int_{t_0}^{t_0 + \Delta t} \vec{v} d t,
\end{equation}

\begin{equation}
\lim _{\Delta t \rightarrow 0} \quad
\overrightarrow{\Delta r}={\vec{v}} \cdot \Delta t, \quad
\vec{r}_{new}^{\prime}=\overrightarrow{r_{old}}+{\vec{v}} \cdot \Delta t
\end{equation}

The simulation was changed from 2D in Python to 3D in Godot.

The stars and the satellite were modeled in Blender and exported in the .obj format.
All textures on the exported models were in the \textit{Principled BDSF} format.

The models were re-textured, and the glow of the stars were re-added with a \textit{WorldEnvironment} node, under the main simulation scene (\textit{System.tscn}).
The corona effect of the stars was added with the help of the PulseGlow shader for Godot (a shader add-on by @fbcosentino). \pagebreak

\begin{figure}[h]
  \centering
  \includegraphics[width=14 cm]{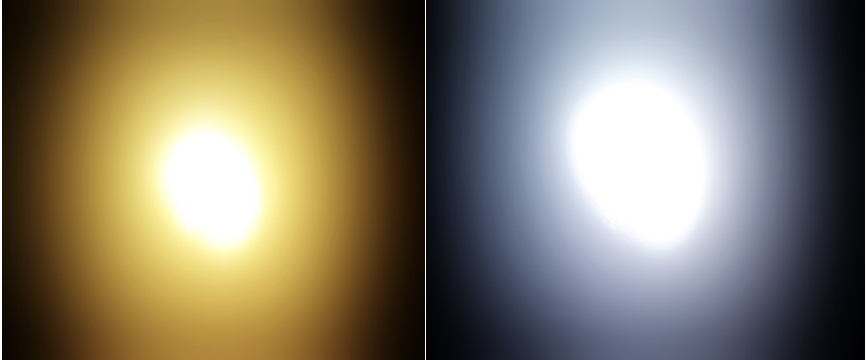}
  \caption{The Stars modeled in Blender.}
\end{figure}

 \begin{figure*}[ht!]
            \centering
            \includegraphics[width=.4\textwidth]{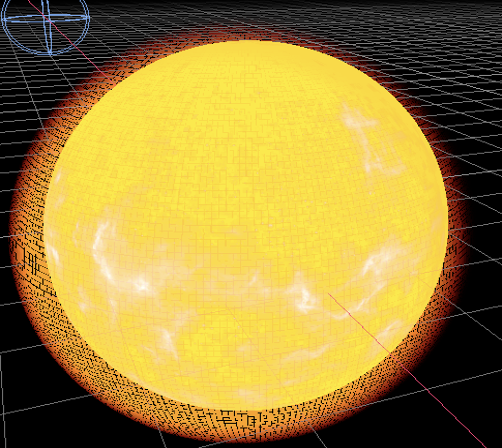}
            \centering
            \includegraphics[width=.416\textwidth]{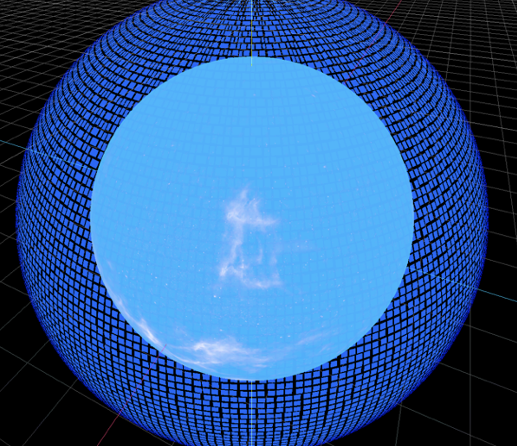}
            \caption{Close-ups of the Godot implemented star models.}
            
        \end{figure*}

\subsection{Scoring Function}

In order to ascertain the relative success of each simulation trial, a scoring function was added onto the final Godot simulation. The scoring function adds the increment value (one minus the ratio of the distance to the closest star over 1000) in each delta time frame, where delta is a parameter of the \textit{process} procedure of the Godot game engine. If the satellite “has made observations of both stars” (has entered within a radius of 200 of each star), the overall score is multiplied by two. The scoring function terminates at the end of each simulation run.
\begin{equation}
\textbf{Score}=[(r_1 flag^{\wedge} r_2 flag).2] \cdot
\sum_{i=delta}^{Simulation Time} 1- \frac{r_{close}}{1000}
\end{equation}

\pagebreak

\section{Conclusion}

In conclusion, a realistic and interactive binary star simulation based on classical mechanics was developed.
The simulation aims to provide an intuition of orbital mechanics in the form of a game; to that end, easily navigable menu screens were developed.

 \begin{figure}[h]
  \centering
  \includegraphics[width=16 cm]{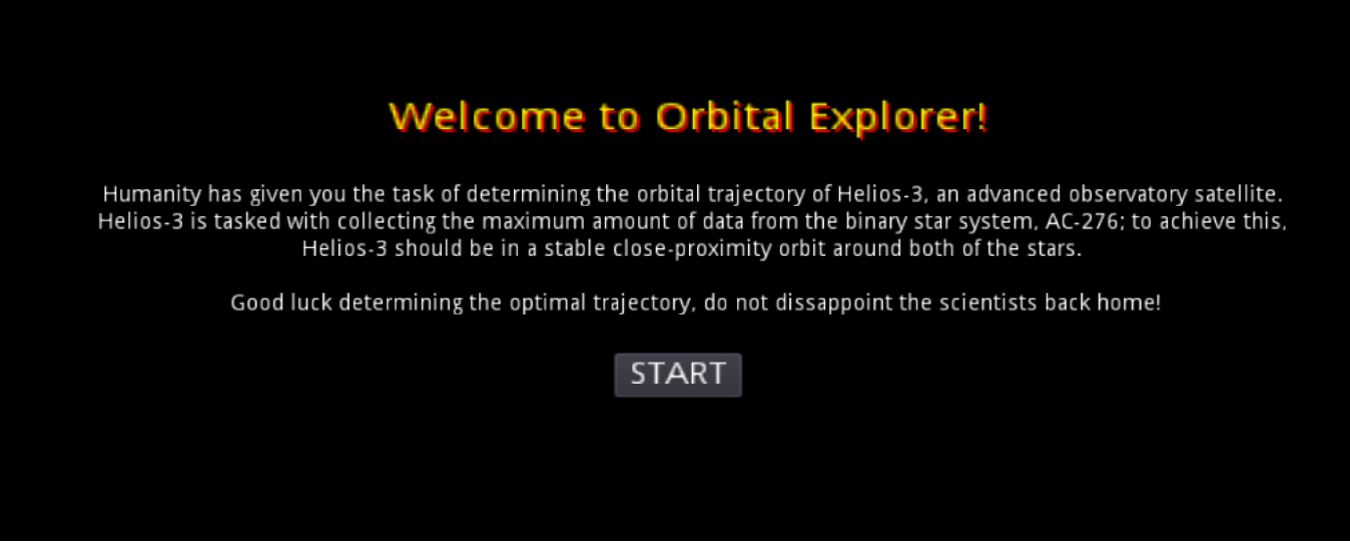}
  \caption{The "Start" Screen}
\end{figure}

 \begin{figure*}[ht!]
            \centering
            \includegraphics[width=.47\textwidth]{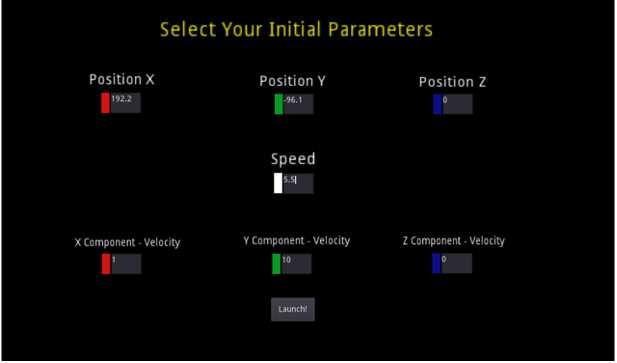}
            \centering
            \includegraphics[width=.52\textwidth]{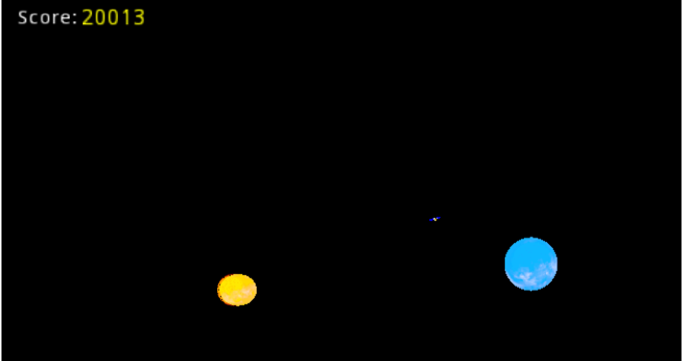}\hfill
            \caption{The "Parameters" Screen and an ongoing simulation.}
            
        \end{figure*}

 \begin{figure*}[ht!]
            \centering
            \includegraphics[width=.475\textwidth]{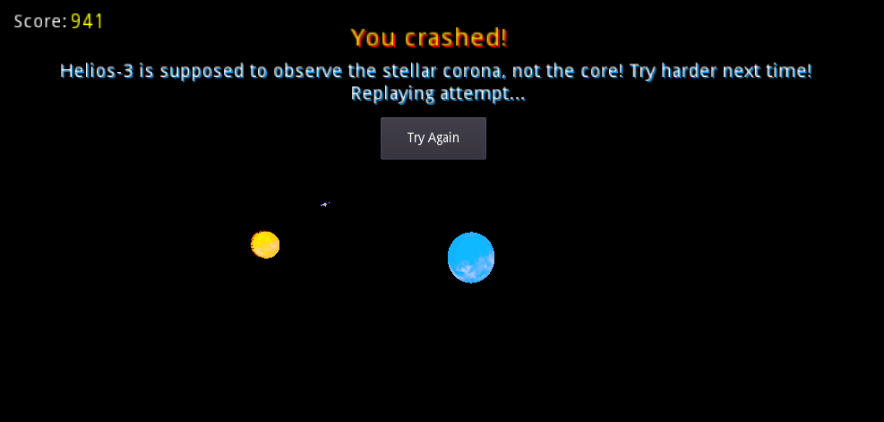}
            \centering
            \includegraphics[width=.52\textwidth]{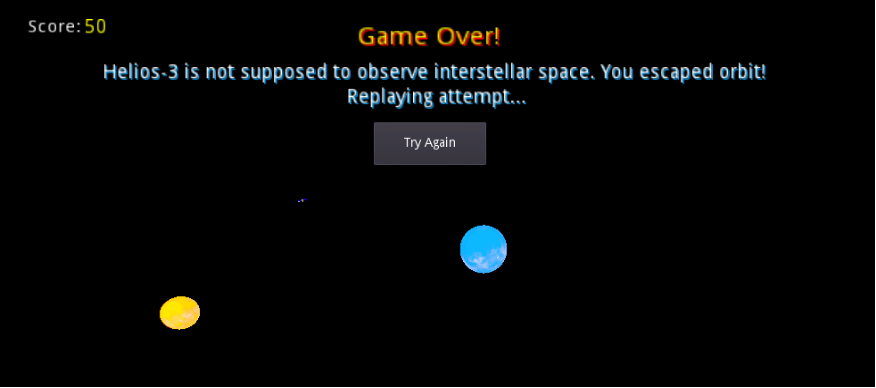}\hfill
            \caption{The “Crash” game-over screen and “Escaped orbit!” game-over screens.}
            
        \end{figure*}

In the simulation, the aim is to get Helios-3, an observatory satellite, in an orbit as close as possible to each of the stars to make the necessary scientific observations. The player provides the initial parameters for the simulation (the velocity and position vectors of the satellite). In each trial, the “usefulness of each run to the scientists” is evaluated with the help of a scoring function. After each run, the player is able to see their score, the replay of their run and why their run has ended.

\bibliographystyle{unsrt}  
\bibliography{references}

\begin{thebibliography}{10}

\bibitem{newton}
Isaac Newton.
\newblock {\em Philosophiæ Naturalis Principia Mathematica}.
\newblock Londini Apud G., J. Innys Second, 1726.

\bibitem{Whittaker}
Steven~N. Shore.
\newblock Celestial mechanics.
\newblock {\em Encyclopedia of Physical Science and Technology (Third
  Edition)}, pages 527--540, 2003.

\bibitem{deAlmeidaJunior2022}
Allan~Kardec de~Almeida~Junior and Antonio Fernando~Bertachini
  de~Almeida~Prado.
\newblock Comparisons between the circular restricted three-body and
  bi-circular four body problems for transfers between the two smaller
  primaries.
\newblock {\em Scientific Reports}, 12(1):4148, Mar 2022.

\bibitem{webb}
Clampin M. et~al. Gardner~J.P., Mather~J.C.
\newblock The james webb space telescope.
\newblock {\em Space Sci Rev 123}, pages 485–--606, 2006.

\bibitem{Musielak}
:~Z~E Musielak and B~Quarles.
\newblock The three-body problem.
\newblock {\em Rep. Prog. Phys. 77 065901}, 2014.

\bibitem{nasa}
Yvette Smith.
\newblock Discovering circumbinary star systems, 2021.
\newblock [Online; accessed October 2, 2022].

\bibitem{learn}
Wen-Chung Shih, Shian-Shyong Tseng, and Chao-Tung Yang.
\newblock Deployment of interactive games in learning management systems on
  cloud environments for diagnostic assessments.
\newblock In Maiga Chang, Wu-Yuin Hwang, Ming-Puu Chen, and Wolfgang
  M{\"u}ller, editors, {\em Edutainment Technologies. Educational Games and
  Virtual Reality/Augmented Reality Applications}, pages 492--496, Berlin,
  Heidelberg, 2011. Springer Berlin Heidelberg.

\bibitem{learn2}
Kuo-En Chang, Yu-Lung Chen, He-Yan Lin, and Yao-Ting Sung.
\newblock Effects of learning support in simulation-based physics learning.
\newblock {\em Computers \& Education}, 51(4):1486--1498, 2008.

\bibitem{1}
Vladimir~A. Chobotov.
\newblock {\em Orbital Mechanics, Third Edition}.
\newblock American Institute of Aeronautics and Astronautics, Inc., 2002.

\bibitem{2}
Knudsen~J M and Hjorth~P G.
\newblock {\em Elements of Newtonian Mechanics: Including Nonlinear Dynamics}.
\newblock (Heidelberg: Springer), 2000.

\end{thebibliography}

\end{document}